\documentstyle[12pt,aasms4]{article}

\def\etal { et al. }
\def\ie { i.\,e. }

\def \om {\omega}

\def \gm {\gamma}
\def \beq {\begin{equation}}
\def \EEQ {\end{equation}}
\def \ppv {{\partial\over{\partial p_{\perp}}}}
\def \ppp {{\partial\over{\partial p_{\parallel}}}}
\def \ku {k_{\parallel}p{_\parallel}}
\lefthead{Ma et al.}
\righthead{Positron-Cyclotron Maser}

\begin{document}
\title{Positron-Cyclotron Maser for the Core Emissions from Pulsars}

\author{Chun-yu Ma \altaffilmark{1}
~~~Ding-yi Mao \altaffilmark{2}
~~~De-yu Wang \altaffilmark{1} 
~~~Xin-ji Wu \altaffilmark{3}} 
\altaffilmark{1}{Purple Mountain Observatory, Academia Sinica, Nanjing 
210008, P.R. China}
\altaffilmark{2}{Department of Mathematics and Physics, Hehai University, 
               Nanjing, China}
\altaffilmark{3}{Department of Geophysics, Peking University, Beijing 
100871, P. R. China} 
\begin{abstract}
We use the cyclotron-maser theory to explain the core emission from
the magnetosphere of pulsars.
As a kind of direct and efficient maser type of emission, 
it can give rise to escaping radiation with
extremely high brightness temperature and narrow angle with respect to the
magnetic axis. We find that the growth
rates and real frequencies of the O-mode electromagnetic wave
propagating parallel to the magnetic fields 
depend on the ratio of  the plasma
frequency $\omega_p$ and the gyrofrequency $\omega_b$
rather than the plasma frequency
alone, as described by other models. The emission takes place in the
region where the magnitude of $\omega_p/\omega_b$ is $10^{-2}$. 
The corresponding altitude is about a few decades of neutron star
radius, where the magnetic field strength is about $10^6-10^8 G$. 
The qualitative spectrum and the lower frequency cut-off 
of the radio emission is obtained by this model. 
\end{abstract}

\section{Introduction}
There is still no satisfactory model for the radio-emission for
pulsars despite more than two decades of theoretical effort.
The radio-emission mechanisms for pulsars are unclear
 because the plasma parameters in the pulsar magnetosphere are
 poorly known.
 The properties of a plasma in superstrong magnetic field as high as
 $10^{12}G$ on the surface of the neutron star
 is not familiar to us as well.\\
The extremely high brightness temperatures observed from radio
pulsars can only be obtained by the coherent motion of the charged particles
in magnetized plasmas. 
The radiation intensity from coherent emission
is proportional to $N^2$ rather than N (the number of particles) 
as compared with incoherent emission (Melrose 1990).
So the incoherent emissions such as cyclotron radiation, synchrotron radiation
or inverse-Compton scattering should not be regarded as the favored mechanism
for pulsar radio emission.
Several candidates of coherent emission mechanisms
such as coherent curvature radiation,
free electron maser emission,
relativistic plasma emission
and cyclotron maser emission
have been suggested.
The coherent curvature emission was the first one put forwards to explain 
the pulsar radio emission. The most serious
 difficulty lies mainly in that the bunching instability
can not afford any velocity dispersion.
The relativistic plasma emission is not a kind of escaping mode itself. 
A conversion mechanism like nonlinear wave-wave interaction
is necessary to couple the  plasma turbulence into escaping radiation, but 
the conversion process limits the radiation efficiency.
The free electron maser seems attractive primarily because it is a 
direct maser emission. But the weakness is that
a large amplitude electrostatic wave should be excited, which acts as
wiggle field in the free electron laser. The generation of the required
parallel electric fields is not clear and the efficiency decreases strongly
with increasing Lorentz factor of the energetic particles
(Melrose 1986).\\
Cyclotron instability developping near the light cylinder of the
pulsar and other plasma processes, which may be responsible for the
generation of radio emission of pulsars, were investigated by the
Geogian group (Machabeli \& Usov 1989, 1979; Kazbegi \etal 1991). The
source region based on their models seems too further from the star as
compared to the recent observation results, \ie, heights ranging
between 1 and 2\% of the light-cylinder radius
 (Hoensbroech \& Xilouris 1996). \\
The cyclotron maser 
 mechanism has been successfully used to interpret planetary radio 
 emissions (e.g. Aurora Kilometer Radiation from the earth), 
solar decimeter spike bursts,
 and radio emissions from extragalactic jets (Wu 1984; 
 Yoon \& Chang 1989; Ma \& Wang 1995).
 The core emissions of radio pulsars was first studied 
by Zhu\etal (1994) and Wang\etal (1989).
Recently, a number of techniques including pulsar timing, scintillation,
pulsar-width narrowing and polarization were developped to derive emission
heights (Melrose 1996;
Hoensbroech \& Xilouris 1996; 
Kijak \& Gil 1996;
Kramer 1995). 
The emission height of a few decades of a neutron star radius is derived 
for many pulsars. \\
Ginzburg \etal (1975)  first pointed out that 
the high intensity of pulsar radiation might be due to an
amplification by maser outside the source rather than in the source. But no 
such an object has been found by now. According to the cyclotron-maser theory,
the electromagnetic wave is amplified inside the source, and it is a kind of
escaping wave itself, i.e., masing takes place inside the source.\\ 
In this paper, we use cyclotron maser theory in the magnetosphere of the pulsar
and calculate the growth rates of micro-instabilities at different altitudes
in the magnetosphere of neutron stars. We find that the growth rates of the 
instability depends on the ratio between the plasma frequency $\omega_p$ and
the gyrofrequency $\omega_b$ rather than the plasma frequency itself. 
The cyclotron maser emission arises only in the region where the value of
 $\omega_p/\omega_b$ falls between 0.01 and 0.1 if the Lorentz factor
of the superthermal particles is $10^3 - 10^4$.
On the other hand, the ratio $\omega_p/\omega_b$
depends on the altitude above the neutron star. So we can reduce the
emission altitudes according to the regions where the growth rates are
positive. 
The emission altitudes obtained by this model are consistent with the
Radius-to-Frequency Mapping (RFM) observed  recently  by the Effelsberg 100m
radio telescope (Kramer \etal 1996; Kijak \etal 1996;
Hoensbroech \etal 1996). In addition, we can get
the low-frequency cut-off from this model.
\section{The parameters in the magnetosphere of neutron stars} 
The magnitude of the magnetic field on the surface of the neutron star
can be estimated from the observed period derivative as following
(Manchester \& Taylor 1977)
\begin{equation}
B_0=3.2\times 10^{19}(P\dot P)^{1/2}\sim10^{12} G ,
\end{equation}
where $P$ is the rotation period of the neutron star in seconds,
 and $\dot P$ is the time derivative of $P$.
 According to the polar cap model of pulsars, a very strong electric field
 associated with enormous potential drop ($V\simeq 10^{12} - 10^{16} V$)
 is available just above the surface of the neutron star owing to the 
rotating magnetic field. So a double layer appears near the surface of the 
neutron star. The 
 charged particles can be accelerated up to very high energy with 
 Lorentz factor as large as $\gm_0\simeq 10^7 - 10^8$ in the double layer.
When those primary particles move along the magnetic lines, they emit
 gamma photons and lose their energy until Lorentz factor falls to about
 $\gm_p=10^3 - 10^5$ (Beskin \etal 1993). At that moment, the primary 
particles reach the altitude in the magnetosphere of few decades 
of neutron star radius where the cyclotron maser growth peaks.   
It is these superthermal primary particles which excite the cyclotron
maser instability in the magnetosphere of the pulsar.
The energy of the gamma photons emitted by the energetic primary particles
is high enough to decay into pairs as they propagate across
 the magnetic field lines. We call those secondary particles the 
stream plasma.\\
According to the Sturrock-Ruderman-Sutherland model, 
the stream plasma frequency is (Manchester \& Taylor 1977, Wu \& Chian 1995)
\begin{equation}
{\omega_p(r)\over \gamma_s^{1/2}} = \left({8e\gamma_p\Omega B_0\over m_e c}\right)^{1/2}
\left(R\over r\right)^{3/2} ,
\end{equation}
where $R$ is the neutron star radius, $\gamma_s$ the Lorentz factor of
the stream plasmas, and $\Omega$ is the angular rotation frequence of the 
pulsar. We adopt $\gamma_s=30-100$ in our calculation (Beskin \etal 1993).
For a dipolar magnetic field, the gyrofrequency of the stream 
electrons or positrons is
\begin{equation}
\omega_b(r)={eB_0\over \gamma_sm_e c}\left(R\over r\right)^3 .
\end{equation}
Eq. (2) and (3) are different from the expressions in the book of 
Manchester \& Taylor by the factor of $\gamma_s$, because we consider 
that the background plasma produced by cascade multiplication outside
the double layer is relativistic stream too.  This improvement was also 
made by other authors (Wu \& Chian 1995, Beskin \etal 1993).  The Lorentz
factor of the stream plasma is of order about 100.\\
We can derive the ratio of plasma frequency and cyclotron frequency 
\begin{equation}
{\omega_p(r)\over\omega_b(r)}
=1.7\times 10^{-9}(\gamma_p\gamma_s^3)^{1/2}
(PB_{0 12})^{-1/2}r^{3/2},
\end{equation}
where the magnetic field $B_{0 12}$ in unit of $10^{12} G$ and $r$ in unit of
 the neutron star radius $R$. 
In Sec. 4, We will illustrate that the growth rates strongly 
depend on the parameter ${\om_p(r)/\om_b(r)}$,
which is a function of the radius from the center of the neutron star.
The cyclotron maser can only take place in the region where the 
magnitude of the ratio ${\om_p(r)/\om_b(r)}$ is about 0.01-0.10.
Once the $\omega_p/\omega_b$ is determined from the numerical calculation,
the emission altitudes can be deduced from Eq. (4). We find that the
emission altitudes range in a few decades of neutron star radius where
the magnetic field is about $10^6-10^8 G$.
\section{Positron-cyclotron maser emission}
We consider two components of the plasma in the magnetosphere of
pulsars:
(1) the high energy primary paricles, $\gamma_p=10^3 - 10^5$;
(2) the cold stream plasmas, $\gamma_s=30 - 100$. 
As a beam of energetic positrons streams through the plasma layer
above the polar cap of a pulsar, the electromagnetic waves can be
directly amplified due to wave-particle resonant interaction. It is
illustrated that only the O-mode wave propagating parallel the
ambient magnetic field is a possible candidate for the core emission
(Zhu \etal 1994).
The dispersion relation of the electromagnetic wave 
propagating along the ambient magnetic field is 
\begin{eqnarray}
N^2 & = &
1+{\sum_\alpha}{\omega_{p\alpha}^2\over\omega}\int d^3p
{1\over\gamma_\alpha\omega\pm\omega_b-\ku}
{p_\perp^2\over 2\gamma_\alpha}\times \nonumber\\
& & \left({\gamma_\alpha\omega-\ku\over p_\perp}\ppv+
k_\parallel\ppp\right) F_\alpha(p_\perp,p_\parallel) ,
\end{eqnarray}
where $\alpha$ denotes different components in the plasmas, and
$F_\alpha(p_\perp,p_\parallel)$ is the distribution function.
To sum over $\alpha$ types of
components,\ie primary particles and magnetosphere plasma streams,
we get
\begin{eqnarray}
N^2 & = & 1-{\omega_p^2\over{\omega(\omega\mp\omega_b)}}+
\pi{\omega_{pp}^2\over\omega}
\int_{-\infty}^{\infty} dp_{\parallel}
\int_{0}^{\infty} dp_{\perp}
{1\over{\gamma_p\omega\pm\omega_b-\ku}} \nonumber\\
& &{p_\perp^2\over{\gamma_p}}
\left({\gamma_p\omega-\ku\over p_\perp}\ppv+k_\parallel\ppp
\right) F_p(p_\perp,p_\parallel) ,
\end{eqnarray}
here $\omega_{pp}$ is the plasma frequency of the primary particles.
According to the wave-particle resonant condition 
\begin{equation}
\gamma_p\omega_r-\omega_b-\ku=0,
\end{equation}
We can see that
only the positrons with positive $\omega_b$ have a contribution to
the cyclotron maser emission.
We write the frequency of the excited wave as
$\omega=\omega_r+i\omega_i$, here $\omega_i$ devotes the growth rate of the wave.
The real part $\omega_r$ meets the dispersion equation of the background plasma. 
The dispersion relation of the O-mode wave propagating parallel to the 
ambient magnetic field $B_0$ with frequency 
\begin{equation}
N^2={c^2k^2\over\omega^2}=1-{\omega^2_p\over\omega_r(\omega_r+\omega_b)}
\end{equation}
By use of the resonant condition and the dispersion relation, this yields
\begin{eqnarray}
\om_i& = & 
{\pi^2\over G}{\omega_{pp}^2\over\omega_r}
\int_{-\infty}^{\infty} dp_\parallel
\int_{0}^{\infty}dp_\perp
\delta(\gamma_p\omega-\omega_{b}-\ku)
{p_\perp^3\over\gamma}\times \nonumber\\
& & \nonumber\\
& &\left({\gamma_p\omega-\ku\over p_\perp}{\partial\over
\partial p_\perp}+k_\parallel{\partial\over\partial
p_\parallel}\right)
F_p(p_\perp,p_\parallel) ,
\end{eqnarray}
where
$$G=2-{\omega_p^2\omega_b\over\omega_r(\omega_r+\omega_b)^2}$$
we consider the primary positrons to be a hollow-beam distribution
\begin{eqnarray}
F_p(p_\perp,p_\parallel)&=&{exp(-\xi^2_0)\over{1+\xi_{0}Z(\xi_{0})}}
(\pi\alpha^2_\perp)^{-1}(\pi\alpha^2_\parallel)^{-1/2}\nonumber\\
& & \nonumber\\
& & exp\left[
-{(p_\perp-p_{0\perp})^2\over{\alpha_\perp^2}}
-{(p_\parallel-p_{0\parallel})^2\over{\alpha_\parallel^2}}
\right]
\end{eqnarray}
where $\xi_0=-ip_{0\perp}/\alpha_\perp$, $\alpha_\perp$ and $\alpha_\parallel$
 denote the characteristic velocity spreads, and 
the average momenta of primary particles are
\begin{equation}
{p_{0\perp}\over mc}\simeq\sqrt{\gamma_p\left(2+\gamma_p\right)}\sin{\psi ,}
\end{equation}
\begin{equation}
{p_{0\parallel}\over
mc}\simeq\sqrt{\gamma_p\left(2+\gamma_p\right)}\cos{\psi ,}
\end{equation}
where $\psi$ is the pitch angle of the particles. It must be pointed
out that the energetic particles lose their momentum component
perpendicular to the magnetic field because of synchrotron losses. 
But the energy of the cyclotron maser comes from the perpendicular 
momentum of the primary positrons. 
As a result, the energetic particles acquire non-zero pitch angle. 
But the quasi-linear wave-particle interaction brings about the diffusion
of particles in the momentum space both along and across the magnetic
field. The waves, which are not escaping mode in the plasmas,
may be excited by cyclotron instability, beam instability \etal 
(Lominadze \etal 1983).
On the other hand, considering the initial energy of the primary 
particles $\gamma_0\sim 10^7 -10^8$, their average pitch angle
$\psi=5^\circ$ and the magnetic field on the surface of the neutron star
$B_0=10^{12}G$, traversing one scale height of the magnetosphere
$\sim r$, we get $r/R>100$ when the final energy is $\gamma\sim 10^4$.
Hence, in order for the model to be plausible, the wave particle 
interaction should be effectice over a distance smaller than a scale 
height and occur exactly where the maser growth rate peaks. We adopt a 
finite value of pitch angle $\psi=5^\circ$ in our calculations.
Finally the  growth rate is written as
\begin{eqnarray}
{n_s\over{n_p}}{\omega_i\over{\omega_b}} & = &
-{2\sqrt{\pi}A\over{G}}
\left(\omega_p\over{\omega_b}\right)^2
\left(\omega_r\over{\omega_b}\right)^{-2}
\int_{p_{-}}^{p_{+}}dp_\parallel \bar p_\perp \times\nonumber \\
& & \nonumber\\
& & \left[
{\bar p_{\perp}-p_{0\perp}\over{\alpha_\perp^2}}
+N\left({\om_r\over{\omega_b}}\right){\bar p_\perp\over C}
{p_\perp-p_{0\parallel}\over\alpha_{\parallel}^2}
\right]\nonumber\\
& & \nonumber\\
& & exp\left[
{(\bar p_{\perp}-p_{0\perp})^2\over{\alpha_\perp^2}}-
{(p_\parallel-p_{0\parallel})^2\over{\alpha_\parallel^2}}\right]
\end{eqnarray}
where $n_s$ and $n_p$ are the densities of the stream plasma and the primary
particles respectively.
$$A={C^2\over\alpha^2_\perp\alpha_\parallel}
{exp(p^2_{0\perp}/\alpha^2_\perp)\over 1+\sqrt{\pi}(p_{0\perp}/\alpha_\perp)
[1+erf(p_{0\perp}/\alpha_{\perp})]exp(p^2_{0\perp}/\alpha^2_\perp)}$$
$erf$ is the error function, and
$$\bar{p}_\perp=C\sqrt{(N^2-1)(p^2_\parallel/C^2)+2N(\omega_r/\omega_b)^{-1}
(p_\parallel/C)+(\omega_r/\omega_b)^{-2}-1}$$
%
The lower and upper limits of the integral are\\
$$p_\pm=C\left[{N\over 1-N^2}\left({\omega_r\over\omega_b}\right)^{-1}\pm
{1\over 1-N^2}\sqrt{N^2-1+(\omega_r/\omega_b)^{-2}}\right]$$
\\
In the next section, we will use Eq. (13) to calculate
the growth rates numerically and derive the radiation altitudes. 
\section{Numerical results}
\begin{figure}
\plotone{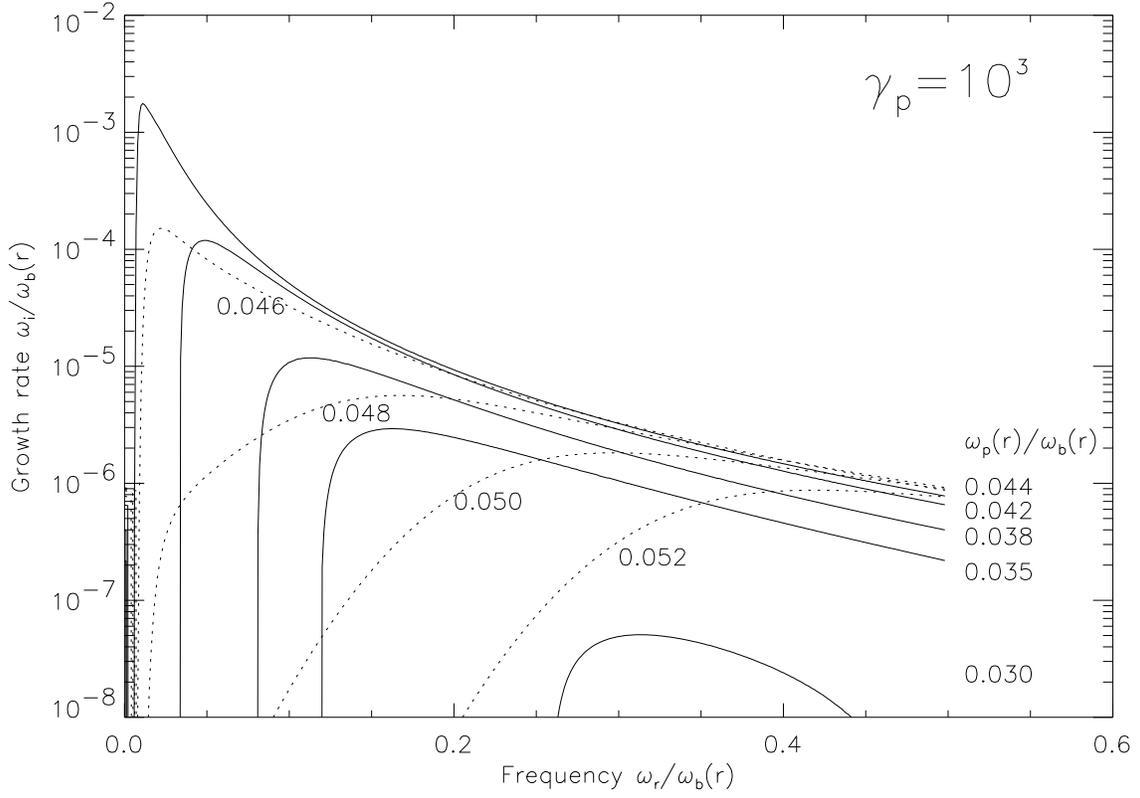}
\caption{The growth rates $(n_p/n_b)(\omega_i/\omega_b)(r)$
         {\sl vs.} the frequencies $\omega_r/\omega_b(r)$. 
         for different ratios of
         plasma frequency and gyrofrequency. All the frequencies are
         normalized by the local gyrofrequency. The Lorentz factor of the
         primary particles is $10^3$. 
       \label{fig1}
}
\end{figure}

\begin{figure}
\plotone{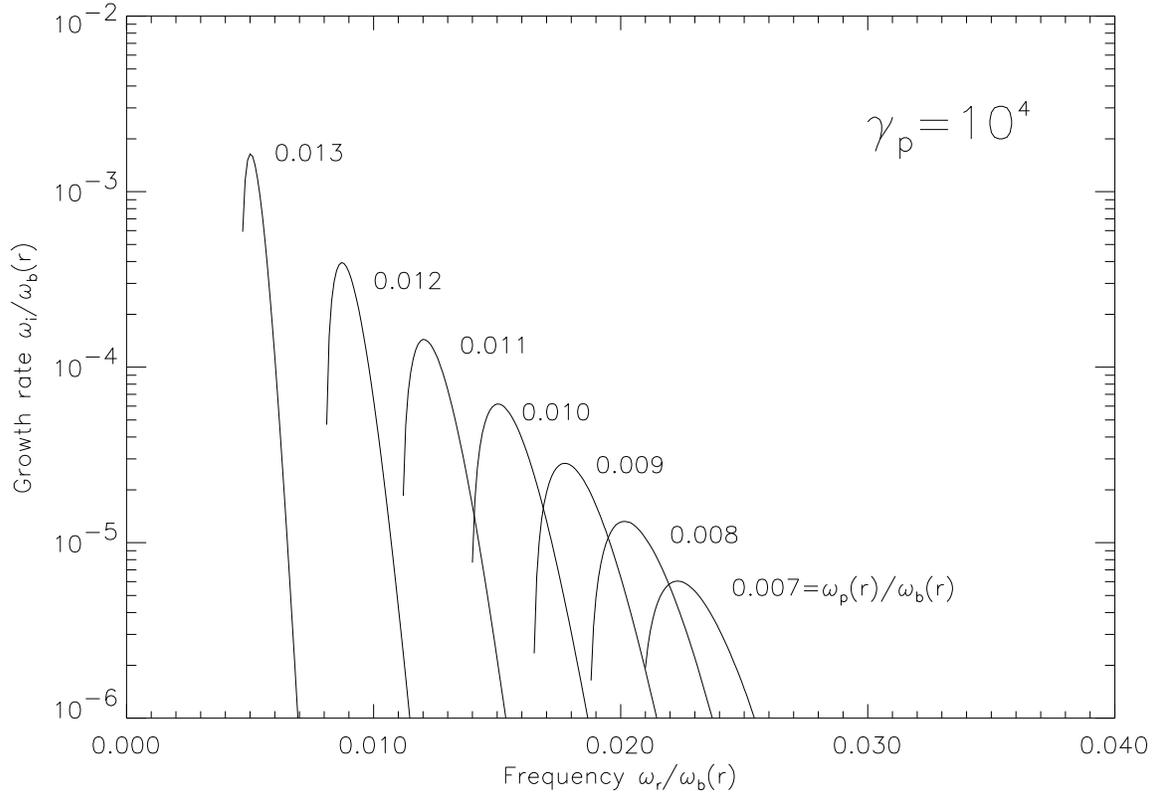}
\caption{The growth rates $(n_p/n_b)(\omega_i/\omega_b)(r)$
         {\sl vs.} the frequencies $\omega_r/\omega_b(r)$. 
         for different ratios of
         plasma frequency and gyrofrequency. 
         The Lorentz factor of the primary particles is $10^4$. 
       \label{fig2}
}

\end{figure}

We set the momentum spreads of the primary particles 
 $\alpha_\parallel/mc=\alpha_\perp/mc=50$, and the pitch
angle of the primary particles $\psi=5^\circ$.  
The energies of the primary particles equal
 $\gamma_p=10^3$ and $10^4$ in the calculations respectively.
The pitch angle can be even smaller although it will decrease the
growth rates of the amplified waves.
The results are displayed in Fig. 1 and Fig. 2.\\
In Fig.1, we show the growth rates of the cyclotron maser instability
for different values of 
$\omega_p/\omega_b$ as
 $\gamma_p=10^3$.
When the ratios of $\omega_p/\omega_b$. 
increases from 0.03 to 0.044, the growth rates increase 
while the real frequencies decrease continuously. If the ratios of
$\omega_p/\omega_b$ increase further, the growth rates decrease but the
real frequencies increase. The growth rates reach a maxium value at
 $\omega_p/\omega_b=0.044$, which corresponds the peak of the
radiation spectrum. So The spectral peak lies at about $\sim 0.01\omega_b$
according to Fig.1 and Fig. 2.,
a lower frequency cut-off appears beyond the turn-over point.
It is of interesting that
the frequency decreases at first, then increases beyond some point as
the altitude rises. In fact, in contrast to most pulsars,
PSR B0525+21 and PSR B0740-28 emit the higher frequencies in
higher emission altitudes (Hoensbroech \& Xilouris 1997).\\
\placefigure{fig1}
In Fig.2, all the parameters are the same as in Fig.1 except for
$\gm_p=10^4$. The significant emissions arise in the region where 
the ratios of $\omega_p/\omega_b$ fall in [0.010, 0.013].
We can see that the frequency decreases monotonously with radius 
in this case. When $\omega_p/\omega_b=0.013$, the growth rate gets its maxium. 
If we adopt $\omega_p/\omega_b=0.014$, the growth rate is almost zero
at any real frequency, \ie, the frequency spectrum is cut-off suddenly. 
Because $\omega_p/\omega_b\propto r^{3/2}$, we can see from Fig. 2 that
the lower frequencies arise from higher altitudes in the magnetosphere of
neutron star in case when the Lorentz factor of the energetic particles
is $10^4$. 
herefore, we can conclude that the Lorentz factors of primary particles
in the emission regions are larger than $10^3$ for most pulsars.
\placefigure{fig2}
It should be mentioned
that all the frequencies are normalized by local gyrofrequency which is
a function of the radius $r$. 
From Eq. (4) we can derive the absolute emission altitudes 
according to the ratios $\omega_p/\omega_b$ with positive growth rates.
For a comparison to our results, we analyzed three samples, \ie, 
PSR 0833+45, PSR 1133+16, and PSR 1929+10.
In Table 1, we reduce the emission altitudes of the three pulsars in case of
$\gm_p=10^3$ and $10^4$ respectively.

\placetable{table-1}
It is inspiring to compare our results with the observed results by 
Radius-to-Frequency Mapping (Kramer 1995, Melrose 1996). 
In Table 2 we display the emission altitudes $r$ in unit of the neutron
star radius $R$,
frequencies $f$ in GHz, and growth rates in $n_p/n_s s^{-1}$ 
for three pulsars when $\gm_p=10^3$.
We can see that the emission is created within a very compact region.
The radius to frequency mapping obtained by this model is much steeper than 
observed. For example, from Table 2, we find that changing the radius
by a factor of $\sim 10\%$ leads to a change in frequency by a factor of 
$\sim 40$. It can probably be improved when the nonlinear saturate processes
of the instability are considered.
The altitude is about a few decades of neutron star radius which is 
qualitatively coincident with the oberved results.  
If $\gamma_p=10^4$, the emission should be closer to the surface of the
neutron star, and the emission frequency decreases monotonously as the
height increases. Therefore, we can predict that the Lorentz factor of primary
particles should be larger than $10^3$ if the observed frequency fellows
the power law $\omega_r\sim r^\nu$ 
(Hoensbroech \& Xilouris 1996, Kijak \& Gil 1996, Kramer 1995). 
\placetable{table-2}
\section{Conclusion remarks}
The emission height is related to the ratio of local plasma frequency
and local gyrofrequency, not to the local plasma frequency alone as
described by other models (Beskin \etal 1993, Melrose 1992).
The Radius-to-Frequency Mapping property exists because the ratio of
local plasma frequency and local gyrofrequency depends on the altitude
above the neutron star.
Significant amplification of escaping radiation takes place in the area
where the ratio $\om_p/\om_b$ is of magnitude about $10^{-2}$ and the 
relevant altitude is a few hundreds kilometers.
Our results are consistent with the Radius-to-Frequency Mapping model and we
can compare them with the observed results
(Hoensbroech \etal 1996, Kijak \etal 1996, Kramer 1995). 
It should be pointed out that the polarization of the O-mode waves 
in our model is circular polarization. However,
if we consider the e$^-$-e$^+$ plasmas in the magnetosphere, the polarization
of normal waves would be linear polarization (Beskin, Gurevich, \&
Istomin 1993).
According to the relation between the growth rate and the real frequency
(see Fig. 2),
we can analyze the emitting spectrum qualitatively with this linear model.  
A lower frequency cut-off is obtained in this model.
To get a quantitative spectral shape, however, we must consider that the 
nonlinear wave-wave interactions in plasma limits the growth of the 
instabilities. We can only predict the spectral index after
considering the nonlinear saturation processes.\\
\acknowledgements
CYM is grateful to R.T. Gangadhara, 
W. Sieber, M. Kramer, and A. Hoensbroech for discussions and comments.
Special acknowledgements are given to Prof. Peter L Biermann for his
hospitality when CYM and DYW had their
visit in Max-Planck Institut F\"ur Radioastronomie.

\clearpage
\begin{table*}
\caption{Emission altitudes $R_{em}$ in unit of neutron star radius
 $R$ of three pulsars for $\gamma=10^3$ and $10^4$ respectively.
\label{table-1}
}
\begin{tabular}{llllllll}
\noalign{\smallskip}
\hline                                                          
\noalign{\smallskip}
Sources & $P$ & $\dot P$ & $B_0$
&\multicolumn{2}{c}{$\omega_p/\omega_b$} 
&\multicolumn{2}{c}{$R_{em}$}\\ 
\noalign{\smallskip}
\cline{5-8}
\noalign{\smallskip}
& $s^{-1}$ &$10^{-15}s^{-1}$&$10^{12}G$ & $\gamma_p=10^3$ &$\gamma_p=10^4$ 
&$\gamma_p=10^3$ &$\gamma_p=10^4$ \\
\noalign{\smallskip}
\hline
\noalign{\smallskip}
PSR 0833+45 & 0.08930 & 124.7  & 1.055 & $1.7 10^{-4}$ 
&5.5 $10^{-4}$ & $41\pm 6$ & 4 - 8\\
PSR 1133+16 & 1.023 & 3.733 & 6.180 & $2.1 10^{-5}$  
&6.8 $10^{-5}$ &$163\pm 15$ & 18 - 33\\
PSR 1929+10 & 0.2265 & 1.157 & 1.619 & 8.8 $10^{-5}$ 
&2.8 $10^{-4}$ & $63\pm 5$ & 7 - 13\\
\noalign{\smallskip}
\hline
\end{tabular}
\end{table*}

\clearpage

\begin{table}
\caption{Emission altitudes $r$ in $R$, 
frequencies $f$ in GHz and the maxium growth rates 
$\omega_i/\omega_b\times (n_s/n_p)$ of the three pulsars in case of
$\gamma_p=10^3$.
\label{table-2}
} 
\begin{tabular}{lllllllll}
\noalign{\smallskip}
\hline                                                          
\noalign{\smallskip}
\multicolumn{3}{c}{PSR0833+45} 
&\multicolumn{3}{c}{PSR1133+16}
&\multicolumn{3}{c}{PSR1929+10}\\
\noalign{\smallskip}
\hline 
\noalign{\smallskip}
 $r$ & $f$ & $\omega_{i max}$ &
 $r$ & $f$ & $\omega_{i max}$ &
 $r$ & $f$ & $\omega_{i max}$ \\
\noalign{\smallskip}
\hline
\noalign{\smallskip}
35 & 1.2  & 1.4 $10^7$ & 139 & 0.11  & 1.4 $10^6$& 54 & 0.30  & 3.5 $10^6$\\
37 & 0.54 & 4.1 $10^7$ & 146 & 0.035 & 4.2 $10^6$& 57 & 0.17  & 1.3 $10^7$\\
39 & 0.22 & 3.5 $10^8$ & 157 & 0.02  & 3.4 $10^7$& 61 & 0.07  & 1.1 $10^8$\\
41 & 0.03 & 4.5 $10^9$ & 162 & 0.003 & 3.6 $10^8$& 62 & 0.01  &1.5 $10^9$\\
42 & 0.07 & 3.6 $10^7$ & 167 & 0.007 & 3.4 $10^6$& 64 & 0.02  &1.1 $10^8$\\
43 & 0.35 & 1.2 $10^7$ & 171 & 0.034 & 1.2 $10^6$& 66 & 0.11  &3.9 $10^7$\\
44 & 1.0  & 3.8 $10^6$ & 176 & 0.095 & 3.6 $10^5$& 68 & 0.30  & 1.1 $10^6$\\
45 & 1.2  & 1.7 $10^6$ & 181 & 0.13  & 1.6 $10^5$& 70 & 0.37  & 5.0 $10^6$\\
\noalign{\smallskip}
\hline
\noalign{\smallskip}
\end{tabular}
\end{table}

\clearpage

\clearpage

\end{document}